\documentclass[
  twocolumn,
  10pt,
  aps,
  prb,
  amsmath,
  amssymb,
  superscriptaddress,
  floatfix
]{revtex4-2}

\usepackage{graphicx}
\usepackage{array}

\usepackage{amsmath,amsfonts,amssymb}
\usepackage{mathtools}
\usepackage{bm}
\usepackage{bbold}
\usepackage{cancel}
\usepackage{mleftright}
\usepackage{comment}
\usepackage{hyperref}

\begin{document}

\title{\texorpdfstring{Interplay between superconductivity and magnetization\\ in strongly coupled disordered SF junctions}{Interplay between superconductivity and magnetization in strongly coupled disordered SF junctions}}

\author{A.\,V.~Levin}
\affiliation{Faculty of Mathematics and Physics, University of Ljubljana, Jadranska 19, 1000 Ljubljana, Slovenia}
\affiliation{Nanocenter CENN, Jamova 39, 1000 Ljubljana, Slovenia}
\author{P.\,M.~Ostrovsky}
\affiliation{Max Planck Institute for Solid State Research, Heisenbergstr.\ 1, 70569 Stuttgart, Germany}

\begin{abstract}
We investigate the superconducting proximity effect in a disordered heterostructure consisting of a bulk superconductor and a thin ferromagnetic layer. Cooper pairs penetrating into the ferromagnet reduce its magnetic stiffness and can drive a transition to a spatially nonuniform magnetic state. Using the semiclassical formalism, we account for the spatial variations of superconducting correlations and determine the effective ferromagnetic stiffness of the SF bilayer. We also argue that the system undergoes a continuous phase transition from the uniform into the helical magnetic state and construct the full phase diagram at arbitrary temperatures. We have solved the general problem numerically and also obtained a number of analytical results in the limits of weak and strong ferromagnets for the equilibrium wave vector of the emergent helical magnetic state.
\end{abstract}

\maketitle

\section{Introduction}

The superconducting proximity effect has attracted considerable interest for more than six decades, dating back to the early years of BCS theory \cite{deGennesGuyon63,deGennes64}. The rich physics of superconductor--normal-metal heterostructures encompasses a wide range of phenomena, including modifications of the local density of states \cite{McMillan68, BelzigBruderSchon96, Devoret96}, Andreev reflection \cite{Andreev64}, and Josephson currents through normal-metal weak links \cite{Kulik70}. An even broader range of phenomena arises in superconductor--ferromagnet heterostructures. The interplay between superconducting correlations and the ferromagnetic exchange field results in the suppression and spatial oscillations of spin-singlet correlations \cite{Buzdin05}, $0$--$\pi$ transitions in Josephson junctions \cite{Ryazanov01}, and long-range spin-triplet correlations capable of carrying supercurrents through ferromagnets \cite{BergeretVolkovEfetov01, BergeretEfetov05, Klapwijk06, Khaire10, Martinez16}. 

Beyond the effects on superconducting correlations mentioned above, superconductivity can modify the magnetic ground state itself. The possibility that superconductivity may favor a nonuniform magnetic configuration over uniform ferromagnetic order was first proposed by Anderson and Suhl \cite{AndrSuhl59}. They argued that superconductivity does not necessarily suppress the ordering of localized magnetic moments but may instead favor a nonuniform magnetic structure, known as the cryptoferromagnetic state. Subsequent experiments on magnetic superconductors revealed the coexistence of superconductivity and spatially modulated magnetic order \cite{Lynn81, Sinha82}, stimulating further theoretical studies of such states \cite{Bulaevskii83}.  In these systems, superconductivity and magnetic ordering coexist within the same material, in contrast to superconductor--ferromagnet heterostructures.

In heterostructures involving both superconducting and normal components, superconducting correlations propagate into the normal parts of the system via the proximity effect. Microscopically, this corresponds to the penetration of coherent Cooper pairs or, equivalently, to Andreev reflection of electrons at the boundaries between different constituent elements of the structure. Electrons in a conventional Cooper pair are arranged in the spin-singlet state, which is incompatible with uniformly ordered ferromagnets. On the one hand, the exchange interaction between itinerant electronic and localized magnetic spins leads to pair-breaking phenomena and suppresses the proximity effect. On the other hand, the same interaction softens the ferromagnetic order and can establish a nonuniform cryptoferromagnetic state.

A microscopic study of the modification of ferromagnetic order induced by the superconducting proximity effect was carried out by Bergeret, Efetov, and Larkin \cite{BergeretEfetov00}.
They considered a thin ferromagnetic layer in contact with a bulk disordered superconductor and investigated the possible formation of a cryptoferromagnetic state with helically rotating magnetization near the critical temperature $T_c$ of the superconductor. The authors demonstrated the possibility of a second-order phase transition between uniform and helical magnetic states. However, their estimates indicated that such a transition was unlikely in the heterostructures then available with strong ferromagnets, such as Fe/Nb \cite{Zabel96, Khaliullin98}.

The influence of superconductivity on magnetic ordering was also investigated by Aikebaier et al. \cite{Aikebaier2019}. They considered a quasi-one-dimensional ferromagnetic-insulator/superconductor structure containing a single N\'eel domain wall. Within the Usadel formalism, they numerically studied the superconducting contribution to the domain-wall energy, taking into account possible spin-orbit scattering and the self-consistent suppression of the superconducting order parameter. They demonstrated that superconductivity reduces the equilibrium domain-wall width and may favor the formation of a dense domain structure for sufficiently weak ferromagnetic stiffness.

Our previous work \cite{LevinOstrovsky26} was dedicated to a tunnel junction between a thin disordered ferromagnet and a bulk clean superconductor (see Fig. \ref{fig:SF}). Treating the ferromagnet as an effectively two-dimensional system and using the Usadel equation, we derived the free-energy functional for an arbitrary magnetic configuration in the vicinity of the phase transition between uniform and nonuniform magnetic orders. This formalism works at all temperatures below $T_c$ and determines the energetically favorable magnetic configuration near the phase transition. For helical magnetic order, a full analytical description, equally valid near and far from the phase transition, was derived.

In the present work, we will investigate superconductivity-induced nonuniform magnetic ordering in a similar SF setup with a diffusive superconductor at arbitrary temperatures below $T_c$. In contrast to Ref.~\cite{BergeretEfetov00}, our approach is not restricted to temperatures close to the superconducting transition. Furthermore, unlike Ref.~\cite{LevinOstrovsky26}, we will consider a junction with strong coupling between the ferromagnet and the superconductor. In the framework of the Usadel equation, we will consider both singlet and triplet correlations induced inside the superconductor. The self-consistent suppression of the superconducting order parameter near the interface (inverse proximity effect) will be also fully taken into account. We will focus on helical magnetic order and determine the equilibrium wave vector over a broad range of exchange fields and temperatures.

The paper is organized as follows. In Sec.\ \ref{Sec:Model of the junction}, we introduce the theoretical framework describing the S/F bilayer. In the following two sections, we develop a convenient parametrization of the quasiclassical Green functions and present a comprehensive numerical solution of the problem with helical magnetization, taking into account the inverse proximity effect. In Sec.\ \ref{Sec:Analytical limits}, we derive analytical results in several limiting regimes, including the cases of weak and strong exchange fields. Finally, we summarize our findings and discuss the relevant experimental setup in Sec.\ \ref{Sec:Summary}. In Appendix \ref{Appendix: gamma}, we numerically investigate the effect of a finite interface resistance and compare our results with those of Ref.\ \cite{LevinOstrovsky26}. Appendix \ref{App:optimal_helix} contains a rigorous proof that the planar helical state is the optimal magnetic configuration.

\section{Model of the junction}
\label{Sec:Model of the junction}

\begin{figure}
    \includegraphics[width=0.7\columnwidth]{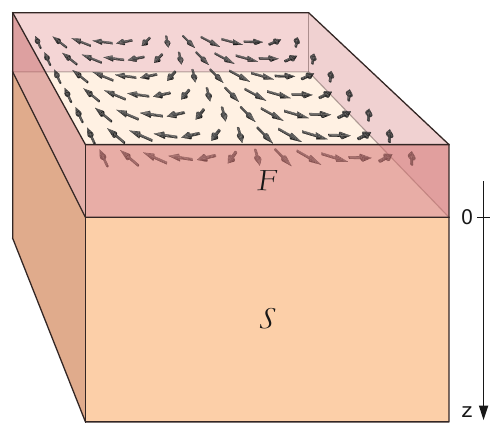}
    \caption{SF junction between a bulk superconductor and a thin ferromagnetic layer. Presence of the superconductor can induce a nonuniform spin texture in the ferromagnet due to direct proximity effect. This results in a spatially varying exchange field $\mathbf{h}(x,y)$.  On the other hand, inverse proximity effect leads to suppression of the order parameter in the superconducting part of the junction near the interface.
}
    \label{fig:SF}
\end{figure}

We consider a hybrid superconductor–ferromagnet (SF) system composed of a bulk superconductor and a relatively thin ferromagnetic layer deposited on its surface (see Fig.\ \ref{fig:SF}). The bulk superconductor is characterized by the value of its order parameter $\Delta_\infty$ far from the surface. Since the system has only one superconductor and carries no equilibrium current, we can always choose $\Delta_\infty$ to be real. Near the SF surface, the value of the order parameter $\Delta(z) < \Delta_\infty$ is suppressed due to the inverse proximity effect but remains real. The ferromagnetic layer can be described by the local exchange field $\mathbf{h}(x,y)$ whose magnitude is fixed while the direction can vary freely along the surface. The thickness $d$ of the ferromagnet is assumed small enough to neglect any dependence of $\mathbf{h}$ on $z$.

We will describe the whole SF junction by its free energy $F$, which consists of three separate contributions:
\begin{equation}
 F = F_\text{mag} + F_\text{el} + F_\Delta.
 \label{F}
\end{equation}
The first contribution, $F_\text{mag}$, is the intrinsic magnetic energy of the ferromagnetic layer, the term $F_\text{el}$ is responsible for the electron dynamics both in the superconductor and in the ferromagnetic layer, and $F_\Delta$ is the free energy associated with the superconducting condensate. 

We adopt a minimal model of the ferromagnet with the energy
\begin{equation}
 F_\text{mag} = \frac{\zeta d}{2} \int d^2 r\: (\nabla \mathbf{n})^2.
 \label{Fmag}
\end{equation}
Here $\zeta$ is the standard bulk magnetic stiffness constant. The thickness $d$ is explicitly separated from $\zeta$ to highlight the scaling of the free energy with the layer thickness. The term $(\nabla \mathbf{n})^2 = (\nabla_x \mathbf{n})^2 + (\nabla_y \mathbf{n})^2$ represents the squared gradient of the magnetization direction, with $\mathbf{n} = \mathbf{h}/h$ being the unit vector along the local exchange field. The form of the magnetic free energy (\ref{Fmag}) is universal and simply represents the leading term in the gradient expansion of the true free energy, which can be any functional of $\mathbf{h}$. To simplify the analysis, we also neglect possible magnetic anisotropy.

The second term, $F_\text{el}$, in Eq.\ (\ref{F}) describes the electron subsystem. We will use the quasiclassical approximation and the limit of relatively strong disorder, so that electron propagation in both the superconductor and ferromagnet is diffusive. Under these assumptions, electronic states can be efficiently described by the quasiclassical Green function $g(\mathbf{r})$. This Green function depends on a single spatial point $\mathbf{r}$ and also acts in the Nambu-Gor'kov and spin spaces, and hence is a matrix of size $4 \times 4$. Within the standard quasiclassical theory \cite{Eilenberger68, Larkin69, BelzigReview99}, this Green function is subject to the normalization condition $g^2 = 1$ and obeys the Usadel equation \cite{Usadel70}. We will use an equivalent approach and work with the quasiclassical action \cite{Altland98} instead,
\begin{equation}
 S[g]
  = \pi \nu \int d^3 r \operatorname{tr} \left[
      \frac{D}{4} (\nabla g)^2 - (\epsilon \tau_z + i \mathbf{h} \bm{\sigma} \tau_z + \Delta \tau_x)g
    \right]. \label{Usadel action}
\end{equation}
Here $\bm{\tau}$ and $\bm{\sigma}$ denote the sets of Pauli matrices in Nambu-Gor'kov and spin spaces, respectively; $D$ is the classical diffusion constant; $\nu$ is the electron density of states and $\Delta(\mathbf{r})$ is the local value of the superconducting order parameter. Solving the Usadel equation for $g$ is equivalent to finding the minimum of the quasiclassical action with respect to $g$. Let us point out that the action contains both the exchange field $\mathbf{h}$ and the superconducting order parameter $\Delta$. Each of these two parameters is present only in the corresponding part of the junction. Another important assumption of our model is a relatively strong coupling between the superconducting and the ferromagnetic parts of the junction (exact conditions are discussed in Appendix \ref{Appendix: gamma}). Hence we assume $g$ to be continuous across the SF interface.

The free energy of the electron subsystem is a sum over fermionic Matsubara energies $\epsilon = (2n + 1) \pi T$ of the above action:
\begin{equation}
 F_\text{el} = T \sum_{\epsilon > 0} S[\epsilon, \Delta, g].
\end{equation}
It is important that the sum here runs over positive Matsubara energies only. This is related to the superconducting symmetry of the underlying Bogoliubov Hamiltonian. In the following, we will mostly consider the zero-temperature limit and replace the above Matsubara sum by an equivalent integral.

The last part of the free energy in Eq.\ (\ref{F}) describes the Cooper interaction between electrons and is responsible for the emergence of superconductivity:
\begin{equation}
    F_\Delta = \nu \int d^3 r \:
    \frac{\Delta^2(\mathbf{r})}{\lambda}.
    \label{F delta}
\end{equation}
Here $\lambda$ is the dimensionless interaction constant. The Cooper interaction is present only inside the superconductor, so the integral here also covers only the superconducting part of the junction. Taking the minimum of $F_\text{el} + F_\Delta$ with respect to $\Delta$ reproduces the standard self-consistent gap equation of the BCS theory.

The complete free energy in Eq.\ (\ref{F}) should be understood as a functional of $\mathbf{h}$, $g$, and $\Delta$. Minimizing it with respect to all three parameters yields a complete solution of the quasiclassical problem. In particular, the spatial dependence of $\mathbf{h}$ and $\Delta$ will provide the optimal magnetization pattern and the gap-suppression profile due to the direct and inverse proximity effects, respectively.

\section{Usadel action}
\label{Sec:Usadel action}

The overall expression for the free energy (\ref{F}) discussed in the previous section is very general and can be applied to a junction of arbitrary configuration. In this section, we will take advantage of the particular SF junction geometry, Fig.\ \ref{fig:SF}, and reduce the Usadel action (\ref{Usadel action}) to a simpler form.

First, we separate the contributions originating from the ferromagnetic layer and the superconductor. Note that the superconducting region contains no exchange field, while the superconducting order parameter $\Delta$ vanishes inside the ferromagnet at $z < 0$. We will focus on the limit of a thin ferromagnetic layer, $d \ll \sqrt{D/h}$. This allows us to neglect possible variations of $g$ inside the ferromagnet along the $z$ direction and effectively treat this part of the junction as two-dimensional. As we have already mentioned earlier, the SF boundary at $z = 0$ is supposed to be sufficiently transparent that the Green function $g$ is continuous across the interface. Exact bounds on the interface conductance and possible effects of an imperfect boundary  are discussed in Appendix \ref{Appendix: gamma}.

In most practically available metallic ferromagnets the exchange field is large compared to the gap in common superconductors. We will hence always assume the limit $h \gg \Delta$ and neglect both the $\epsilon$ and $\Delta$ terms in the action inside the ferromagnet. Within this approximation, we can rewrite the full Usadel action as a sum of the ferromagnet and the superconductor contributions:
\begin{subequations}
\label{S} 
\begin{gather}
 S[g] = \pi \nu_F d \int d^2 r \: L_\text{FM}(g) + \pi \nu_S \int \limits_{z>0} d^3 r \: L_\text{SC}(g),\\
 L_\text{SC}[g] =  \operatorname{tr} \left[ \frac{D}{4} (\nabla g)^2 - (\epsilon \tau_z + \Delta \tau_x)g \right], \label{SSC}\\
  L_\text{FM}(g) =  -i\operatorname{tr} \Bigl[ \mathbf{h} \bm{\sigma} \tau_z \:g \Bigr]_{z=0}. \label{SFM}
\end{gather}
\end{subequations}
Here we have allowed for two different densities of states $\nu_{F/S}$ in the ferromagnet and the superconductor. The last ferromagnetic term in the action effectively sets boundary conditions for $g$ in the superconductor at $z = 0$. 

We can estimate the relative contributions of the ferromagnetic and superconducting regions to the total Usadel action. For the ferromagnet, the characteristic magnitude of the action is $\sim \nu_F h d$ while for the superconductor it is $ \sim \nu_S \Delta \xi$, where $\xi = \sqrt{D/2\Delta}$ is the superconducting coherence length. We introduce a special proximity energy scale
\begin{equation}
    \varsigma = \frac{\nu_S \sqrt{D \Delta /2}}{\nu_F d} 
    \label{varsigma}
\end{equation}
to quantify the relative strength of superconductivity and ferromagnetism in the junction. The ratio $\varsigma/h$ determines the qualitative nature of the proximity effect. In the limit $h \gg \varsigma$, the exchange field dominates and all superconducting correlations inside the ferromagnet are strongly suppressed. In the opposite case $h \ll \varsigma$, the electrons inside the ferromagnet are only weakly affected by the exchange field and the entire SF system behaves effectively as a singlet superconductor. The most interesting interplay between superconductivity and magnetism occurs when both scales are of the same order, $h \sim \varsigma$.

The full quasiclassical matrix Green function is of the size $4 \times 4$ and obeys the constraint $g^2 = 1$. It contains in total eight parameters as was discussed in detail in Refs.\ \cite{IvanovFominov06, LevinOstrovsky26}. In our case, the junction has only one superconducting terminal, and hence no supercurrents flow in the equilibrium state. This allows us to reduce the number of variables in $g$ from eight down to four. The Usadel action (\ref{S}) explicitly contains only the matrices $\tau_x$ and $\tau_z$ in the Nambu-Gor'kov space; both these matrices anticommute with $\tau_y$. Hence the solution $g$ that minimizes the action should also be a matrix that anticommutes with $\tau_y$. We can parametrize such $g$ explicitly as
\begin{equation}
 g
  = \frac{1}{2} \Bigl[ (\tau_z - i \tau_x) Q + (\tau_z + i \tau_x) Q^{-1} \Bigr],
 \label{gQ}
\end{equation}
Here we introduce a new matrix $Q$ that has the size $2 \times 2$ and acts in the spin space only. The condition $g^2 = 1$ is fulfilled automatically, and $Q$ is any invertible matrix with four parameters.

The superconducting part of the action (\ref{SSC}) is fully uniform in the $xy$ plane. The only quantity that varies in these directions is the magnetization $\mathbf{h}$; it enters our model through the boundary term (\ref{SFM}) only. We will perform a gauge transformation of the matrix $Q$ to remove this explicit spatial dependence. Let us define a unitary matrix $W$ acting in the spin space such that
\begin{equation}
 \mathbf{h} \bm{\sigma} = W^{-1} h \sigma_z W.
 \label{hW}
\end{equation}
This rotation simply chooses a local basis in spin space such that, at each point, the $z$ axis is aligned with $\mathbf{h}$. We can now perform a gauge transformation and apply the same rotation to the matrix $Q$:
\begin{equation}
   Q \mapsto W^{-1} Q W.
\end{equation}
In terms of the rotated matrix $Q$, the Usadel action (\ref{S}) becomes
\begin{subequations}
\label{SQ}
\begin{gather}
 \begin{multlined}[b][0.85\linewidth]
 L_\text{SC}[Q] = \operatorname{tr} \biggl[
      \frac{D}{2} (\mathcal{D}  Q) (\mathcal{D} Q^{-1}) \\
      - (\epsilon - i\Delta) Q - (\epsilon  + i\Delta) Q^{-1}
    \biggr],
 \end{multlined} \\
 L_\text{FM}[Q] = -i h \operatorname{tr} \Bigl[
       \sigma_z (Q+Q^{-1})
    \Bigr]_{z=0}.
\end{gather}
\end{subequations}
We have thus eliminated any inhomogeneity in the ferromagnetic part of the action. Instead, the diffusion term in the superconductor now contains long derivatives involving the non-Abelian gauge potential
\begin{equation}
 \nabla \mapsto \mathcal{D} = \nabla + i [\mathbf{A}, \cdot],
 \qquad
 \mathbf{A} = i (\nabla W) W^{-1}.
\end{equation}
The gauge field $\mathbf{A}$ is a vector in real space and a traceless matrix in spin space. It encapsulates all the information about spatial inhomogeneity of the magnetization. Since the magnetization in the ferromagnet changes only in the $xy$ plane, we have $A_z = 0$, so the $z$ component of the long derivative remains unchanged: $\mathcal{D}_z = \nabla_z$.

The choice of the matrix $W$ satisfying Eq. (\ref{hW}) is not unique. We can include any additional rotations in $W$ that preserve the direction of the $z$ axis. Such additional rotations, which have no observable consequences, are pure gauge transformations. They can alter the local value of the vector $\mathbf{A}$. Naturally, our theory of the proximity effect will make only predictions that are insensitive to the choice of the gauge. At the same time, we can use the gauge freedom to simplify some intermediate equations and make the derivation of the final results easier.

\section{Helical state}
\label{Sec:Helical state}

\begin{figure}
    \centering
    \includegraphics[width=\linewidth]{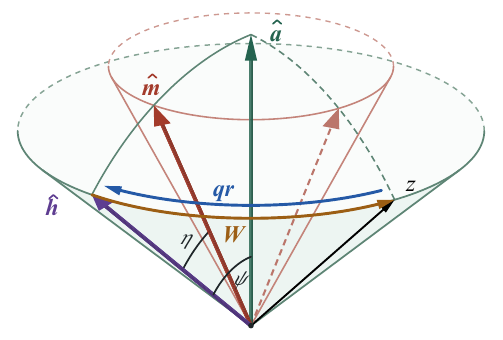}
    \caption{Visualization of the helical state. Magnetization $\mathbf{h}$ rotates in spin space along a cone with the axis $\mathbf{\hat a}$ and half opening angle $\psi$. After the gauge transformation (\ref{hW}), (\ref{Ahelical}), direction of $h$ is fixed along the $z$ axis. Triplet component of the Green function is characterized by the vector $\mathbf{m}$ that lies between $\mathbf{h}$ and $\mathbf{\hat a}$. The angle $\eta$ between $\mathbf{h}$ and $\mathbf{\hat{m}}$ as well as the magnitude of $\mathbf{m}$ changes as a function of $z$ to minimize the Usadel action (\ref{L general a}). For the planar helical state, $\mathbf{h}$ rotates within a single plane ($\psi = \pi/2$) and the vector $\mathbf{m}$ is always parallel to $\mathbf{h}$ ($\eta = 0$).}
    \label{fig:cone}
\end{figure}

For the rest of this work, we will study a specific magnetic texture that is known as a helical state. In this state, the magnetization vector $\mathbf{h}$ rotates in space at a constant rate. In the most general form of the helical state, the magnetization rotates around a fixed direction $\mathbf{\hat a}$ in spin space, tracing a cone with the half opening angle $\psi$, see Fig.\ \ref{fig:cone}. Another characteristic of the state is the wave vector $\mathbf{q}$ in real space that points in the direction along which the magnetization changes and whose magnitude corresponds to the rate of rotation. We can always choose the basis in spin space such that the $z$ axis lies on the surface of the cone. Then the whole helical state can be described by Eq.\ (\ref{hW}) with the following simple rotation matrix:
\begin{equation}
 W = e^{i(\mathbf{q} \mathbf{r}) (\mathbf{\hat{a}}\bm{\sigma})/2},
 \qquad
 \mathbf{A} = - \frac{\mathbf{q}}{2} (\mathbf{\hat{a}}\bm{\sigma}).
 \label{Ahelical}
\end{equation}
The main technical advantage of a generic helical state is that it can always be described by a constant vector potential.

In the most general form, the matrix field Q can be parametrized by the scalar variable $\theta$ and the vector variable $\mathbf{m}$ \cite{IvanovFominov06}:
\begin{equation}
    Q(z) = e^{i \theta + \mathbf{m} \bm{\sigma}}.
    \label{Qq parametrization}
\end{equation}
These variables have a clear physical meaning: $\theta$ is the standard Usadel angle in the spin-singlet channel. Deviations of $\theta$ from zero describe the onset of superconducting correlations. The components of the vector $\mathbf{m}$ have the same meaning but for the three spin-triplet amplitudes of the superconducting correlations.

For a helical state with the constant vector potential (\ref{Ahelical}), the action (\ref{SQ}) is completely uniform in the $xy$ plane. Hence, the optimal matrix $Q$ depends only on $z$. We rewrite the action (\ref{SQ}) in terms of $\theta$ and $\mathbf{m}$ and use the dot symbol to denote the derivative in $z$:
\begin{subequations}
\label{L general a}
    \begin{gather}
\begin{multlined}[b]
 L_\text{SC} = D \left[\dot \theta^2 - \dot m^2
        - \left( {\mathbf{\dot{\hat m}}}^2 + q^2 [{\mathbf{\hat m}} \times \mathbf{\hat{a}}]^2 \right) \sinh^2 m \right]\\
        - 4(\epsilon \cos \theta + \Delta \sin \theta)\cosh m,
\end{multlined} \label{LSC} \\
 L_\text{FM} = 4 (\mathbf{h} {\mathbf{\hat m}}_0) \sin \theta_0 \sinh m_0.
\end{gather}
\end{subequations}
Here we have also used separate notations $m$ and $\mathbf{\hat m}$ for the absolute value and the direction of the vector $\mathbf{m}$. Index $0$ refers to the values of the fields at $z = 0$.

In the uniform magnetic state with $q = 0$, minimum of the action is achieved with the triplet vector $\mathbf{m}$ parallel to the magnetization $\mathbf{h}$. This solution indicates that only one triplet component of the superconducting correlations (the one with zero spin projection on the direction of $\mathbf{h}$) is induced by proximity to a uniform ferromagnet. For a helical state with $q \neq 0$, the minimum of the action (\ref{L general a}) is achieved by some $z$-dependent Usadel angle $\theta$ and the triplet vector $\mathbf{m}$ that lies in the plane of $\mathbf{h}$ and $\mathbf{\hat{a}}$, see Fig.\ \ref{fig:cone}. It turns out that the optimal magnetic configuration, that minimizes the total free energy of the SF junction, always corresponds to $\psi = \pi/2$. We provide a proof of this statement in Appendix \ref{App:optimal_helix}. Magnetization rotates within a single plane in the spin space, hence such a configuration is known as the planar helical state.

The action (\ref{L general a}) greatly simplifies in this case. The direction of the triplet vector $\hat{\mathbf{m}}$ is constant everywhere in the bulk of the superconductor and coincides with the direction of $\mathbf{h}$ in the ferromagnet. It remains to find only the optimal profiles of $\theta$ and $m$. The corresponding equations are

\begin{subequations}
\label{bulk theta-m eqs}
\begin{align}
    \frac{D\ddot\theta}{2} &= \left(\epsilon \sin\theta - \Delta \cos\theta \right) \cosh m, \label{bulk_theta} \\
    \frac{D\ddot m}{2} &= \left( \epsilon \cos\theta  +\Delta \sin\theta + \frac{Dq^2}{2} \cosh m \right) \sinh m.
     \label{bulk_m}
\end{align}
\end{subequations}
They are supplemented by the boundary conditions at the S/F interface ($z = 0$):
\begin{subequations}
\label{boundary theta-m}
\begin{align}
    \dot\theta_0 - \frac{h}{\varsigma \xi} \cos \theta_0 \sinh m_0 &= 0, \label{boundary theta} \\
    \dot m_0 +\frac{h}{\varsigma \xi} \sin \theta_0 \cosh m_0 &= 0. \label{boundary m}
\end{align}
\end{subequations}

These equations are difficult to solve analytically since they contain the order parameter $\Delta$, which is itself  an unknown function of $z$ due to the inverse proximity effect. We should determine $\Delta$ self-consistently together with $\theta$ and $m$. The gap equation for $\Delta$ can be derived by minimizing the free energy (\ref{F delta}) together with (\ref{L general a}):
\begin{equation}
    \Delta_z = 2\pi \lambda T \sum \limits_{\epsilon>0} \sin \theta_z \cosh m_z.
\end{equation}
At large energies, the right-hand side of this equation has the asymptotic behavior $\propto \Delta/\epsilon$, resulting in a logarithmically divergent Matsubara sum. We can regularize this divergence using the standard subtraction procedure: we divide both sides of the gap equation by $\Delta_z$ and then subtract the same equation for large values of $z$, where the inverse proximity effect is negligible. This procedure eliminates both the logarithmic divergence and the coupling constant $\lambda$, giving
\begin{equation}
    \sum \limits_\epsilon \left[ \frac{\Delta_z}{\sqrt{\epsilon^2 + \Delta_\infty^2}} - \sin \theta_z \cosh m_z \right] = 0. \label{sce}
\end{equation}
Here $\Delta_\infty$ denotes the bulk value of the BCS order parameter far from the interface with the ferromagnet. The value of $\Delta_\infty$ changes only with temperature.

We thus have the full set of equations (\ref{bulk theta-m eqs}) and (\ref{sce}) to determine $\theta$, $m$, and $\Delta$ for a given helical state with the wave vector $q$. Once this problem is solved, we can proceed to minimizing the total free energy, including the magnetic stiffness part (\ref{Fmag}), with respect to $q$.

The derivative of the total free energy with respect to $q^2$ is
\begin{equation}
 \frac{\partial F}{\partial q^2}
  = \frac{\zeta - \alpha(q)}{2}.
 \label{Fq}
\end{equation}
The first term represents the magnetic stiffness and originates from Eq.\ (\ref{Fmag}). The second term is due to the electronic parts of the free energy $F_\text{el} + F_\Delta$ and has the form
\begin{equation}
    \alpha(q) = \frac{2\pi \nu_S D T}{d} \sum \limits_{\epsilon>0} \int \limits_0^\infty dz \: \sinh^2 m_z. \label{alpha dev}
\end{equation}
Here $m$ is taken from the solution of Eqs. (\ref{bulk theta-m eqs}), (\ref{sce}), and hence implicitly depends on $q$. The parameter $\alpha(q)$ encapsulates all the information about the proximity effect and quantifies a reduction in the magnetic stiffness of the ferromagnet due to proximity to the superconductor according to Eq.\ (\ref{Fq}). If $\alpha$ is sufficiently large, the proximity effect will overcome the intrinsic stiffness of the ferromagnet. As a result, the ferromagnetic layer will form a helical state with the wave vector $q$ such that $\alpha(q) = \zeta$. 

\begin{figure}
    \centering
    \includegraphics[width=\linewidth]{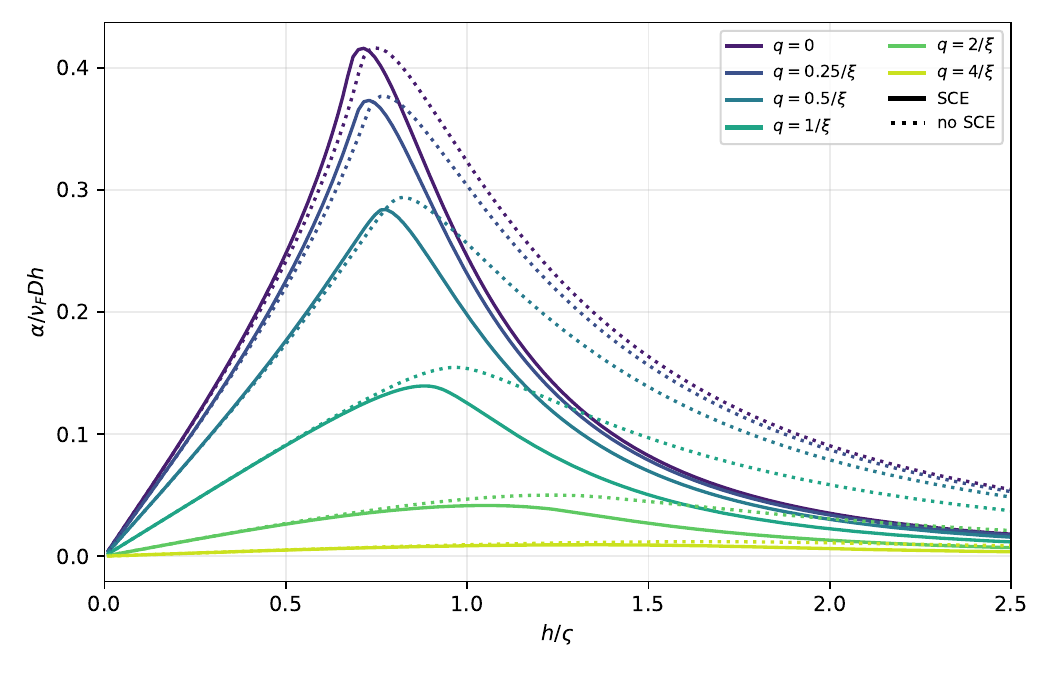}
    \caption{Dependence of $\alpha$ on $h/\varsigma$ for several values of $q$ at $T = 0$. Solid curves are derived from the numerical solution of Eqs.\ (\ref{bulk theta-m eqs})--(\ref{alpha dev}). Colors correspond to different values of $q$ measured relative to the coherence length $\xi = \sqrt{D/2\Delta}$. To illustrate the effect of the self-consistent suppression of the superconducting order parameter (inverse proximity effect), we repeated the calculation without Eq.\ (\ref{sce}), using instead a uniform order parameter $\Delta_z = \Delta_\infty$. These results are shown by the dotted curves.}
    \label{fig:developed helical state}
\end{figure}

\begin{figure}
    \centering
    \includegraphics[width=\linewidth]{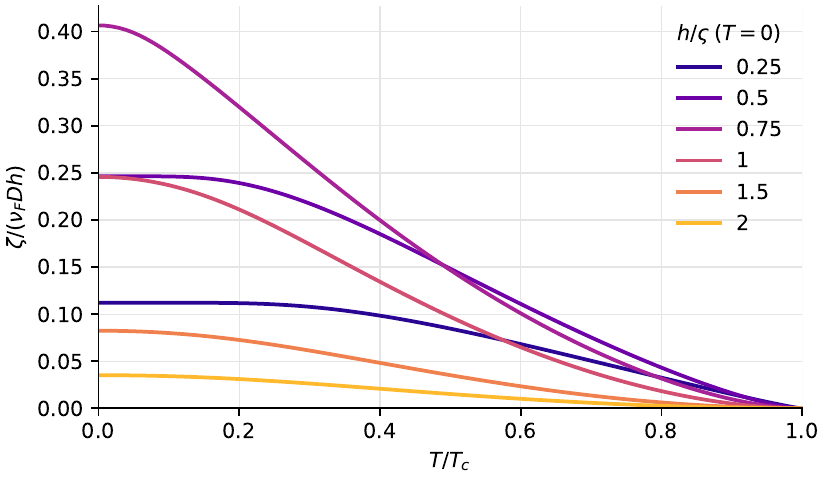}
    \caption{Phase diagram of the transition between uniform and helical states in the $T-\zeta$ plane. The temperature is measured in units of $T_c$. The ferromagnet stiffness $\zeta$ is brought to dimensionless units as $\zeta/\nu_F D h$. Colors correspond to different ratios of the exchange field $h$ and the proximity energy $\varsigma$, defined in Eq.\ (\ref{varsigma}). Each curve separates the uniform (above) and the helical (below) phases of the junction. A second order phase transition between these two states occurs when $\alpha(q=0) = \zeta$, where $\alpha$ is the proximity-induced correction to the magnetic stiffness, see Eq.\ (\ref{alpha dev}).}
    \label{fig:alpha_on_T}
\end{figure}

We have solved the Usadel equations (\ref{bulk theta-m eqs}) together with the self-consistency equation (\ref{sce}) numerically and plot the value of $\alpha$ in Fig.\ \ref{fig:developed helical state} as a function of $h/\varsigma$ for several values of $q$. With increasing $q$, the triplet component $m$ is suppressed, and $\alpha$ decreases. The highest value of $\alpha$ for a given $h/\varsigma$ is attained at $q = 0$. This maximal value determines the position of the second-order phase transition from the uniform magnetic state of the junction to the helical state. We have also plotted the dependence of $\alpha(q=0)$ on temperature in Fig.\ \ref{fig:alpha_on_T}. This plot represents a phase diagram of the junction in the $T$--$\zeta$ plane. 

As was mentioned above, $\alpha(q)$ is a decreasing function of $q$. When $\alpha(q) = \zeta$, the ferromagnet orders in a helical state with the wave vector $q$. At this point, the derivative of the total free energy Eq.\ (\ref{Fq}) vanishes. The second derivative at the same point remains positive:
\begin{equation}
 \frac{\partial^2 F}{(\partial q^2)^2}
  = -\frac{\alpha'(q)}{4q} > 0.
\end{equation}
We thus conclude that the helical state corresponds to a minimum of the total free energy and is hence stable.

We also observe that $\alpha$ has its maximum value when $h \sim \varsigma$, that is, when the typical energy scales of the ferromagnetism and superconductivity are comparable. This is the most favorable condition for inducing helical magnetic order by proximity effect. In our earlier work \cite{LevinOstrovsky26} on the tunnel SF junction, a resonant behavior was observed under similar conditions, and the parameter $\alpha$ diverged at $T = 0$ and $q = 0$. In the present case of a transparent SF junction, a sharp divergence is replaced by a smooth maximum.

In both limits, when either superconducting or magnetic energy dominates, $\alpha$ is small, and the system will likely retain the uniform magnetic order. We also see that when the ferromagnet is relatively strong ($h \gtrsim \varsigma$), suppression of $\Delta$ by the inverse proximity effect plays an important role. It leads to a sizable decrease of $\alpha$. In the opposite limit of a relatively strong superconductor ($h \lesssim \varsigma$), the suppression of $\Delta$ near the interface is negligible and can be disregarded. We will discuss these two limits in more detail below.

\section{Analytical limits}
\label{Sec:Analytical limits}

In the previous Section, we have established the very general equations to determine a possible helical magnetic state in the SF junction. For a given value of the wave vector $q$, one should solve Eqs.\ (\ref{bulk theta-m eqs}) and (\ref{sce}) with the boundary conditions (\ref{boundary theta-m}) to determine $\theta$, $m$, and $\Delta$ as functions of $z$. Then it only remains to calculate $\alpha(q)$ from Eq.\ (\ref{alpha dev}) to find the dependence of the free energy (\ref{Fq}) of the junction on $q$.

We will now consider two major limiting cases of a relatively large or small exchange field $h$. In both limits, the triplet amplitude $m$ turns out to be small and the Usadel equation can be linearized in $m$. As a result, the equation for the singlet parameter $\theta$ and the self-consistency equation for $\Delta$ do not involve $m$. The full set of equations has the form
\begin{gather}
 \frac{D\ddot\theta}{2} = \left(\epsilon \sin\theta - \Delta \cos\theta \right), \label{lin bulk theta} \\
 \frac{D\ddot m}{2} = \left( \epsilon \cos\theta  +\Delta \sin\theta + \frac{Dq^2}{2} \right) m, \label{lin bulk m} \\
 \sum \limits_\epsilon \left[ \frac{\Delta}{\sqrt{\epsilon^2 + \Delta_\infty^2}} - \sin \theta \right] = 0. \label{lin bulk Delta}
\end{gather}

The parameter $h$ enters these equations only through the boundary conditions (\ref{boundary theta-m}). We will analyze these boundary conditions and adjust them to the linearized equations separately for the cases for weak and strong exchange field $h$ below.

Once the linearized equation for $m$ is solved, we can readily find $\alpha(q)$ from the simplified version of Eq.\ (\ref{alpha dev}):
\begin{equation}
    \alpha(q) = \frac{2\pi \nu_S D T}{d} \sum \limits_{\epsilon>0} \int \limits_0^\infty dz \: m^2_z.
    \label{lin alpha dev}
\end{equation}

\subsection{Weak exchange field}

For a relatively weak ferromagnet, $h \lesssim \varsigma$, the inverse proximity effect is negligible. This can be seen from Fig.\ \ref{fig:developed helical state}, where solid and dotted curves (with and without inverse proximity effect, respectively) almost coincide at small $h$. In this regime, the ferromagnet only weakly perturbs the superconducting part of the junction. We may therefore neglect deviations of the singlet angle $\theta$ and of the order parameter $\Delta$ from their bulk values $\theta_\infty$ and $\Delta_\infty$.

The same conclusion can be also drawn from the boundary conditions for $\theta$. Simply neglecting $h$ in Eq.\ (\ref{boundary theta}) reduces it to $\dot\theta_0 = 0$. As a result, Eq.\ (\ref{lin bulk theta}) admits a constant solution $\theta_z = \theta_\infty$ and the self-consistency equation (\ref{lin bulk Delta}) is also automatically satisfied with $\Delta_z = \Delta_\infty$. 

It only remains to solve the linearized Usadel equation for $m$. Since the angle $\theta$ is constant in space, Eq.\ (\ref{lin bulk m}) is solved by the exponential function
\begin{equation}
   m = m_0 e^{-\kappa z}, \quad \kappa^2 = \frac{2}{D}\sqrt{\epsilon^2 + \Delta_\infty^2} + q^2.
\end{equation}
To determine the boundary value $m_0$, we apply Eq.\ (\ref{boundary m}) with the known value of $\theta$:
\begin{equation}
 \dot m_0
  = -\kappa m_0
  = -\frac{h}{\varsigma \xi} \frac{\Delta_\infty}{\sqrt{\epsilon^2 + \Delta_\infty^2}}.
\end{equation}

Finally, substituting this solution for $m$ into Eq.\ (\ref{lin alpha dev}) gives
\begin{equation}
    \alpha(q) = \frac{\pi \nu_F D h^2}{\varsigma} \sum_{\epsilon>0}\: \frac{\Delta_\infty^{5/2}}{(\epsilon^2 + \Delta_\infty^2) \bigl( \sqrt{\epsilon^2 + \Delta^2_\infty} + D q^2/2 \bigr)^{3/2}}.
    \label{alpha small h}
\end{equation}
For fixed material parameters, the right-hand side depends only on the wave vector $q$ and temperature $T$. It decreases with increasing $q$, reflecting the suppression of the triplet component by rapid spatial variations of the magnetization.

The maximal superconducting correction to the magnetic stiffness is reached at $q=0$ and $T=0$. At this point, the parameter $\alpha$ takes the value
\begin{equation}
    \alpha(q = 0) = \frac{\sqrt{\pi} \: \Gamma \left( 1/4 \right)}{12\: \Gamma \left( 3/4 \right)} \frac{\nu_F D h^2}{\varsigma}  \approx 0.437 \frac{\nu_F D h^2}{\varsigma}.
    \label{alpha max small h}
\end{equation}
If the ferromagnetic stiffness $\zeta$ exceeds this value, the SF junction will always remain in the uniform magnetic state. For smaller values of $\zeta$, a transition to the helical state becomes possible.

We can also estimate the parameter $\alpha$ at relatively large $q$ from Eq.\ (\ref{alpha small h}). Summation over Matsubara energies then yields
\begin{equation}
    \alpha(q \gg 1/ \xi) = \frac{\pi}{4}\frac{\nu_F D h^2}{\xi^3 q^3 \varsigma}  \tanh{\frac{\Delta_\infty}{2T}}.
    \label{alpha small h large q}
\end{equation}
When the ferromagnetic stiffness $\zeta$ is small, the helical state with a relatively large $q \sim \zeta^{-1/3}$ emerges. In this limit, rapid spatial variations of the magnetization effectively reduce the value of the exchange field $h$. Hence the weak-ferromagnetic limit (\ref{alpha small h}) actually applies up to $h \lesssim \varsigma \sqrt{q\xi}$. We will derive this upper bound for $h$ in the next subsection.

\subsection{Strong exchange field}

Consider now the opposite limit of a relatively strong exchange field $h \gg \varsigma$. In this regime, the superconducting correlations (both singlet and triplet) are strongly suppressed near the S/F interface. The boundary conditions (\ref{boundary theta-m}) 
can be simplified to the following form:
\begin{equation}
    \theta_0 = 0, \qquad m_0 = \frac{\varsigma \xi}{h}\, \dot \theta_0.
    \label{boundary large h}
\end{equation}
The equations for $\theta$ and $\Delta$ fully decouple from the linearized equation (\ref{lin bulk m}) for $m$ and can be solved independently. The pair of equations (\ref{lin bulk theta}) and (\ref{lin bulk Delta}) is a standard set of the singlet Usadel equation and the self-consistency equation that describes the inverse proximity effect with the full suppression of the superconducting order at the interface of a superconductor and a strong normal metal. This combination of a nonlinear differential equation and an integral equation can only be tackled numerically.

Once this singlet background is determined, it remains to solve the linearized equation (\ref{lin bulk m}) for $m$ and calculate $\alpha$ from Eq.\ (\ref{lin alpha dev}). The magnitude of $m$ is fixed by the boundary condition (\ref{boundary large h}) and scales as $m \sim \varsigma/h$. The numerical solution of the inverse proximity effect problem then yields the following result:
\begin{equation}
    \alpha(q = 0) = 0.265 \frac{\nu_F D \varsigma^3}{h^2}.
    \label{alpha big h}
\end{equation}
This is the maximal possible value of $\alpha$ attained at zero $q$ and $T$.

With increasing $q$, the numerical factor in Eq.\ (\ref{alpha big h}) decreases. Eventually, in the limit of large $q$, the problem simplifies and admits a full analytical solution. At large $q$, the term with $q^2$ dominates in Eq.\ (\ref{lin bulk m}). The solution is then given by a simple exponential function
\begin{equation}
    m_z = m_0 e^{-q z} = \frac{\varsigma \xi}{h}\, \dot \theta_0 e^{-qz}.
    \label{m(z) developed}
\end{equation}
Here we have already used the boundary condition (\ref{boundary large h}) to fix the amplitude of $m$. The parameter $\alpha$ is then given by Eq.\ (\ref{lin alpha dev}) and takes the form
\begin{equation}
    \alpha (q \gg 1/\xi) =  \frac{\nu_F \xi^3 \varsigma^3}{q h^2} 2\pi T \sum_{\epsilon>0}  \dot \theta_0^2.
    \label{alpha dot theta}
\end{equation}
It remains to calculate the Matsubara sum involving $\dot\theta_0^2$.

Remarkably, an explicit solution of Eq.\ (\ref{lin bulk theta}) is not required to evaluate the remaining sum. Instead, using translational invariance of the Usadel equation in the $z$ direction, we can establish the following conserved quantity:
\begin{equation}
    I_z = \sum_{\epsilon>0} \left[\frac{D \dot\theta^2}{4} + \epsilon (\cos \theta - 1) + \frac{\Delta \sin \theta}{2} \right], \quad \dot I = 0.
\end{equation}
The fact that $I$ is independent of $z$ can be easily checked by taking a derivative of the above equation and then applying Eqs.\ (\ref{lin bulk theta}) and (\ref{lin bulk Delta}). At the SF interface, the Usadel angle $\theta$ vanishes [see Eq.\ (\ref{boundary large h})], and hence $I_0$ reduces to exactly the Matsubara sum from Eq.\ (\ref{alpha dot theta}). At the same time, far in the bulk of the superconductor, $\dot\theta$ vanishes, and we can compute $I_\infty$ using the standard bulk Usadel angle $\theta_\infty$. Equating these two values of $I$, we obtain
\begin{equation}
    \frac{D}{4} \sum_{\epsilon>0} \dot \theta_0^2 = \sum_{\epsilon>0} \left[\frac{\epsilon^2 + \Delta_\infty^2/2}{\sqrt{\epsilon^2 + \Delta^2_\infty}} - \epsilon \right].
\end{equation}
This sum can now be inserted directly in Eq.\ (\ref{alpha dot theta}) to provide the final result for $\alpha(q)$ in the limit of large $q$. At zero temperature, it becomes
\begin{equation}
    \alpha(q \gg 1/\xi) = \frac{\nu_F D \varsigma^3}{4q\xi h^2}.
    \label{alpha large h large q}
\end{equation}
The wave vector $q$ therefore scales as $q\propto\zeta^{-1}$ in this regime.

Finally, let us compare the values of $\alpha$ in the limit of large $q$ for relatively weak (\ref{alpha small h large q}) and strong (\ref{alpha large h large q}) exchange fields. These two results match when $\xi q \sim h^2 / \varsigma^2$. We can therefore conclude that the crossover between the limits of weak and strong magnetization is shifted towards larger $h$ as $q$ increases. This can be also seen directly in Fig.\ \ref{fig:developed helical state}: the maximum of $\alpha$ shifts to larger $h$ with increasing $q$.

\section{Summary and discussion}
\label{Sec:Summary}

In conclusion, we have investigated the influence of the superconducting proximity effect on the stability of magnetic ordering in S/F heterostructures consisting of a thin ferromagnetic layer coupled to a bulk superconductor, see Fig.\ \ref{fig:SF}. Using the Usadel equation formalism for the bulk disordered superconductor, we have accounted for the induced spin-triplet correlations and the self-consistent suppression of the superconducting order parameter (inverse proximity effect). The central result is the superconductivity-induced correction $\alpha$ to the ferromagnetic stiffness $\zeta$, given by Eq.\ (\ref{alpha dev}), which applies at arbitrary temperatures below $T_c$ and for any ratio of the exchange field $h$ to the proximity energy $\varsigma$, see Eq.\ (\ref{varsigma}).

We have solved the coupled Usadel and self-consistency equations (\ref{bulk theta-m eqs}) and (\ref{sce}) numerically for a specific magnetic configuration: the planar helical state with the wave vector $q$. Based on this solution, we have plotted the dependence of $\alpha$ on $h/\varsigma$ in Fig.\ \ref{fig:developed helical state} for several values of $q$ at $T = 0$. Another view of the same result is the dependence of $\alpha(q = 0)$ on temperature shown in Fig.\ \ref{fig:alpha_on_T}. The latter plot represents a phase diagram for the transition between uniform and helical magnetic states in the $T$--$\zeta$ plane. It turns out that the proximity correction $\alpha$ to the magnetic stiffness is strongest, and hence the helical phase is the most extensive, when $h \sim \varsigma$. For both large and small values of the exchange field $h$, the value of $\alpha$ decays, and the helical phase shrinks. We have found analytic expressions for $\alpha$ in these two limiting cases, see Eqs.\ (\ref{alpha max small h}), (\ref{alpha small h large q}), (\ref{alpha big h}), and (\ref{alpha large h large q}).

Let us estimate the parameters of a typical experimental setup and assess the possibility of realizing a proximity-induced helical magnetic state. Superconducting heterostructures are commonly based on Nb with the order parameter $\Delta = 17$\,K. We take the coherence length to be $\xi\sim15$\,nm, which is usual for disordered Nb and corresponds to $D_S = 10\, \text{cm}^2/\text{s}$. Assuming comparable densities of states, $\nu_S\sim\nu_F$, and taking the thickness of the ferromagnetic layer $d=5$\,nm, we have the characteristic proximity energy
\begin{equation}
    \varsigma \sim \frac{\Delta \xi}{d} \sim  50 \:\text{K}.
\end{equation}
According to our results, the proximity-induced helical state will most likely appear when the ferromagnet has a comparable magnitude of the exchange field. In the experiments \cite{Ryazanov01} on SFS $\pi$ junctions, the $\text{Cu}_{0.48}\text{Ni}_{0.52}$ alloy with the Curie temperature of approximately $20$--$30$\,K was used. This value is close to the above estimate of $\varsigma$, hence this type of a ferromagnet should be also very well suited for observing the emergent helical state.

In most metallic ferromagnets, the stiffness, the density of states, and the magnitude of the exchange field are related. The following ratio of these quantities has the meaning of the spin-wave stiffness \cite{LandauLifshitz9} and typically falls into a relatively narrow interval \cite{KuzminSkokov20}:
\begin{equation}
    \frac{\zeta}{\nu_F h} \sim  [0.01 - 0.2] \: \text{cm}^2/\text{s}.\label{zeta est}
\end{equation}
Our theory predicts a helical state with a wave vector $q$ such that $\zeta = \alpha(q)$. The above estimate of $\zeta$ implies $\zeta/\nu_F D_S h \sim 0.01$ for Nb. Such a small value of the stiffness should correspond to a relatively large value of $q$. We can estimate it using Eq.\ (\ref{alpha small h large q}):
\begin{equation}
    q \sim 4/\xi \sim 0.3 \: \text{nm}^{-1}. \label{q est}
\end{equation}

For stronger ferromagnets, the value of $q$ gets smaller, and eventually the effect vanishes completely. We can estimate an upper bound on the exchange field with the help of Eq.\ (\ref{alpha big h}):
\begin{equation}
    h \lesssim 3 \varsigma \sim 150 \: \text{K}. \label{h max est}
\end{equation}
This restriction is rather weak; many available ferromagnetic materials are within this bound. We thus believe that experimental observation of helical magnetic order predicted in this paper is feasible.

\begin{figure}
    \includegraphics[width=\columnwidth]{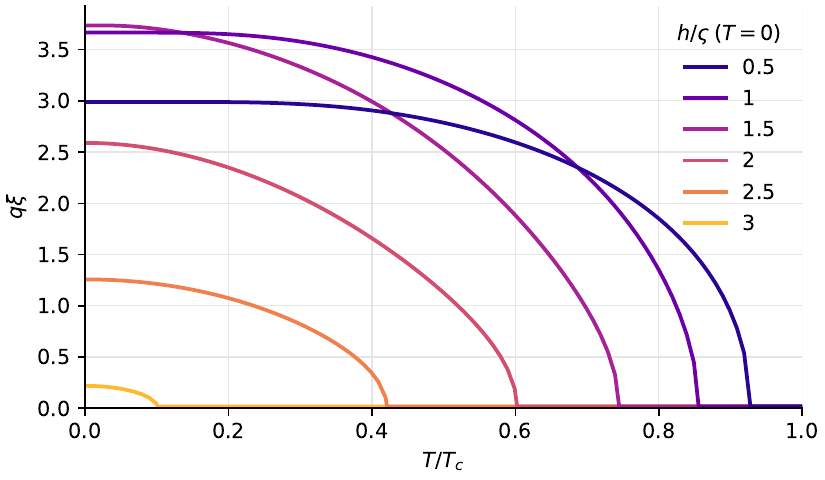}
    \caption{Temperature dependence of the helical wave vector for several values of $h/\varsigma$ and fixed magnetic stiffness $\zeta/\nu_F D h = 0.01$. Each curve shows a second order phase transition from the helical to the uniform magnetic state with increasing temperature.}
    \label{fig::q_T}
\end{figure}

Experimentally, temperature is the most easily controlled parameter. Comparing the ferromagnetic stiffness $\zeta$ with $\alpha(q)$ from Eq.\ (\ref{alpha dev}), we deduce the equilibrium helical wave vector $q$ as a function of temperature, as shown in Fig.\ \ref{fig::q_T}. This dependence has a typical shape characteristic of second-order phase transitions with $q$ vanishing continuously (as a square root) at a certain critical temperature below the superconducting $T_c$. This kind of temperature dependence should be directly observable experimentally. 

In this work, we have used the simplest theoretical model of an isotropic ferromagnet characterized by a single stiffness parameter $\zeta$. More sophisticated models often include possible magnetic anisotropy as well as spin-orbit coupling. These effects can modify the resulting magnetic texture but the main qualitative conclusion of our work remains unchanged: proximity to a superconductor ``softens'' the ferromagnet and can induce a phase transition from the uniform to a nonuniform state.

\acknowledgments{We are grateful to M.\ Feigelman and Ya.\ Fominov for stimulating discussions. The work of AVL was supported by J\v{S}RIPS RS (contract no.\ 11011-3/2024). We also thank A.\ Levina for help with the design of Fig.\ \ref{fig:SF}.}

\appendix

\section{General boundary conditions}
\label{Appendix: gamma}

\begin{figure}
    \includegraphics[width=\columnwidth]{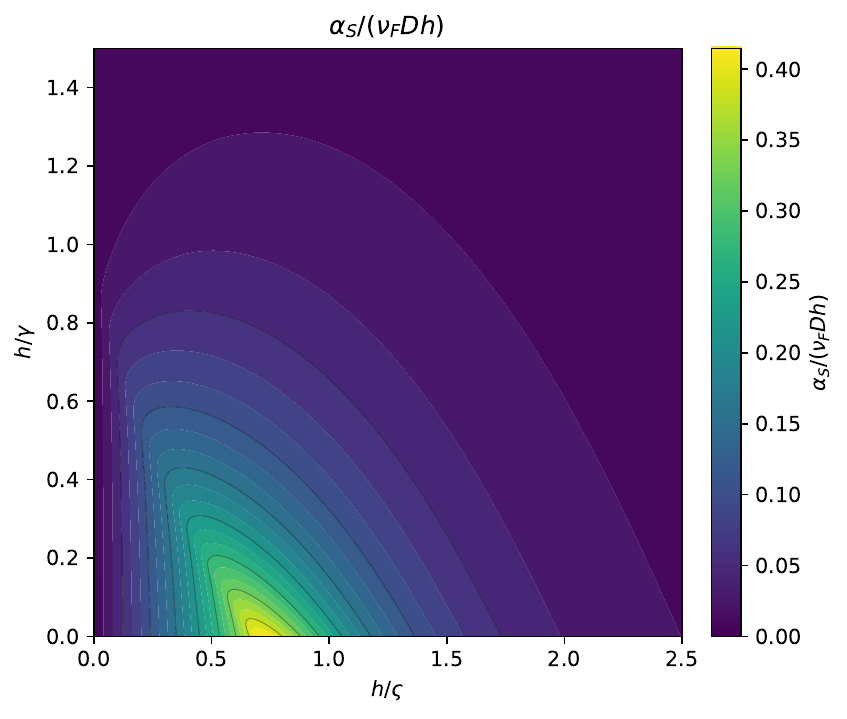}
    \includegraphics[width=\columnwidth]{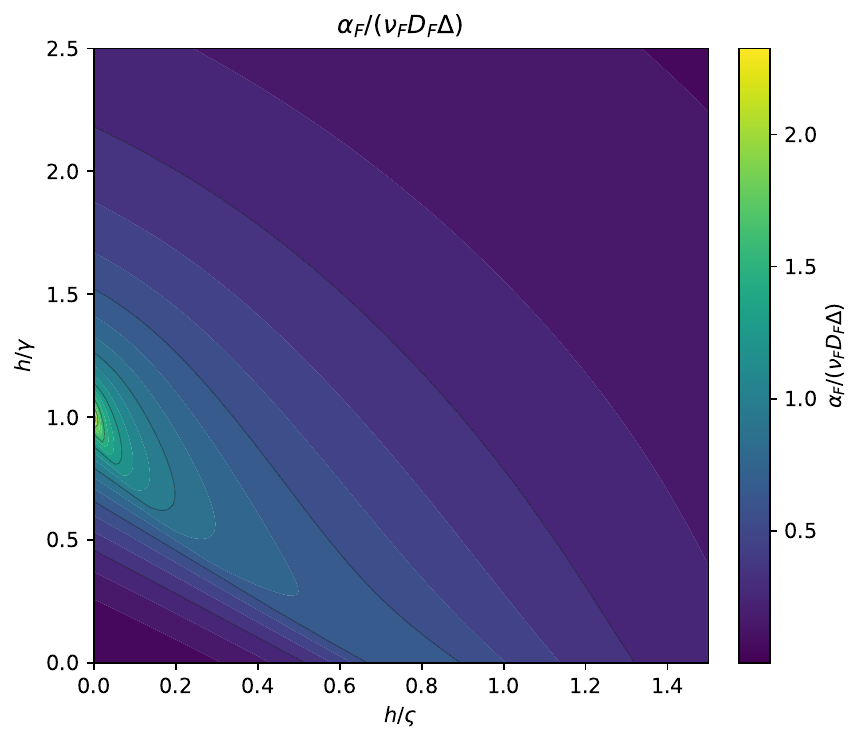}
\caption{Proximity contributions to the magnetic stiffness originating from the superconductor, $\alpha_S$ [Eq.\ (\ref{alphaS}), upper panel], and from the ferromagnet, $\alpha_F$ [Eq.\ (\ref{alphaF}), lower panel], as functions of the exchange field $h$, proximity energy $\varsigma$, and interface coupling energy $\gamma$. For a relatively weak ferromagnet, $h \ll \varsigma$, superconductor contribution $\alpha_S$ is strongly suppressed and only the ferromagnet part $\alpha_F$, which depends on $h/\gamma$, is relevant. This limit was considered in Ref.\ \cite{LevinOstrovsky26}. In the case of strong SF coupling $\gamma \gg h$, main proximity contribution to the stiffness comes from the superconductor, $\alpha_S$, and depends on the ratio $h/\varsigma$. This is the limit studied in the main text of this paper.}
    \label{fig::alpha_gamma}
\end{figure}

In this Appendix, we briefly discuss a more general setup in which the interface between superconducting and ferromagnetic parts of the junction is characterized by an arbitrary conductance $G_T$. In such a case, the semiclassical Green function $g$ is no longer continuous but rather has a finite jump at the boundary. We will retain the notation $g_0$ for the value of the Green function at $z = 0$ on the superconductor's side. In the ferromagnet, the value of the Green function is denoted by $g_F$.

Finite transparency of the SF interface is described by an additional term \cite{KuprianovLukichev88, Nazarov1994} to be added to the Usadel action (\ref{S})
\begin{equation}
 S_B
  = -\frac{\pi G_T}{4} \int d^2 r\,  \operatorname{tr} \bigl[ g_0 \: g_F \bigr],
 \label{SB}
\end{equation}
This boundary action contains an integral over the junction interface. The conductance $G_T$ per unit area is measured in units of $e^2/\hbar$. Let us note that, although the functional form of Eq.\ (\ref{SB}) implies tunneling nature of the boundary, it can be used to describe both weak and strong coupling limits as $G_T$ varies from small to large values.

Similarly to Eq.\ (\ref{varsigma}), we introduce an effective energy scale to describe the coupling between the ferromagnet and the superconductor \cite{LevinOstrovsky26}:
\begin{equation}
    \gamma = \frac{G_T}{4 \nu_F d}.
\end{equation}
Zero value $\gamma = 0$ formally describes a fully uncoupled junction, while $\gamma = \infty$ corresponds to the strong coupling limit when the Green function is continuous through the interface. In the main text of the paper, only this latter limit is discussed. Here, we will consider arbitrary values of $\gamma$ (as compared to $h$ and $\varsigma$) and illustrate the limits of applicability of our main results.

For a finite transparency parameter $\gamma$, boundary conditions (\ref{boundary theta-m}) for the Usadel equations take a more general form:
 \begin{gather}
 \label{dtheta0_general}
    \begin{multlined}[b][0.85\columnwidth]
    \varsigma \xi \dot\theta_0  = \gamma \sin(\theta_0 - \theta_F) \cosh (m_0 - m_F)\\ =  h \cos \theta_F \sinh m_F, 
    \end{multlined}\\
    \label{dm0_general}
    \begin{multlined}[b][0.85\columnwidth]
    \varsigma \xi\dot m_0  =\gamma \cos(\theta_0 - \theta_F) \sinh (m_0 - m_F)\\ = - h \sin \theta_F \cosh m_F. 
    \end{multlined}
\end{gather}
When $\gamma$ is large, these conditions automatically fix $\theta_F = \theta_0$ and $m_F = m_0$, thus recovering Eqs.\ (\ref{boundary theta-m}) from the main text. In the opposite limit of small $\gamma$, the values of $\theta$ and $m$ on both sides of the junction are almost decoupled. As a result, on the superconducting side, we have $\theta_0 \approx \theta_\infty$ and $m_0 \approx 0$. This weak-coupling limit was studied in Ref.\ \cite{LevinOstrovsky26}.

The bulk Usadel equations inside the superconductor and the self-consistency equation for $\Delta$ retain their forms (\ref{bulk theta-m eqs}) and (\ref{sce}). Our main result is formulated in terms of the parameter $\alpha$ that describes the effective suppression of the magnetic stiffness in the ferromagnet (\ref{Fq}) due to proximity to the superconductor. For arbitrary boundary coupling $\gamma$, the parameter $\alpha$ has two contributions, one from each of the ferromagnetic and superconducting parts of the junction:
\begin{subequations}
\label{boundary general}
\begin{gather}
 \alpha = \alpha_F + \alpha_S, \\
    \alpha_F = 2\pi T \nu_F D \sum \limits_{\epsilon>0} \sinh^2 m_F, \label{alphaF} \\
    \alpha_S = \frac{2\pi T \nu_S D}{d} \sum \limits_{\epsilon>0} \int \limits_0^\infty dz \: \sinh^2 m_z. \label{alphaS}
\end{gather}
\end{subequations}
In the limit of strong coupling (large $\gamma$), the second contribution dominates. Hence the definition of $\alpha$ in the main text of the paper, Eq.\ (\ref{alpha dev}), corresponds to $\alpha_S$.

We have solved the set of Usadel equations with the general boundary conditions (\ref{boundary general}) together with the self-consistency equation (\ref{sce}) and identified separately the values of $\alpha_F$ and $\alpha_S$ at $q = 0$ as functions of $h/\varsigma$ and $h/\gamma$. Both quantities are shown in Fig.\ \ref{fig::alpha_gamma}. We observe that in the limit $\gamma \gg h$, the superconducting contribution $\alpha_S$ dominates. This is exactly the limit discussed in the main text of the paper. For a relatively strong superconductor, $\varsigma \gg h$, the main correction to the stiffness comes from the ferromagnet $\alpha \approx \alpha_F$. This is the limit considered in Ref.\ \cite{LevinOstrovsky26}.

\section{Optimality of the planar helical state}
\label{App:optimal_helix}

In this Appendix, we demonstrate that, among conical helical configurations, the planar helical state is always the most energetically favorable.

Magnetization $\mathbf{h}$ rotates along a cone with the axis $\mathbf{\hat a}$ and the half opening angle $\psi$. The triplet vector $\mathbf{m}$ lies in the plane spanned by $\mathbf{h}$ and $\mathbf{\hat{a}}$, see Fig.\ \ref{fig:cone}. The angle $\eta$ between $\mathbf{m}$ and $\mathbf{h}$ is a function of $z$ and $\epsilon$.

Let us introduce the following linear functional that can act on any function $f(z, \epsilon)$:
\begin{equation}
    A[f] = \frac{2\pi\nu_S DT}{d}\sum_{\epsilon > 0} \int \limits_0^\infty dz\, f\,\sinh^2 m.
\end{equation}
In particular, acting on the constant function $f = 1$, the value of this functional reproduces the correction $\alpha$ to the ferromagnetic stiffness, cf.\ Eq.\ (\ref{alpha dev}).

In order to evaluate the optimal configuration of the helical state, we isolate a part of the total free energy that explicitly depends on $q$ and $\psi$. This dependence originates only from the ferromagnetic term (\ref{Fmag}) and from the $q^2$ term in Eq.\ (\ref{LSC}),
\begin{equation}
    F(q, \psi) = \frac{q^2}{2} \left( \zeta\sin^2\psi - A[\sin^2(\psi-\eta)] \right).
    \label{F short}
\end{equation}
Our goal is to minimize the full free energy with respect to both the wave vector $q$ and the half opening angle $\psi$. However, since the full free energy is stationary with respect to the superconducting fields $\theta$, $m$, $\eta$, and $\Delta$, their implicit dependence on $q$ and $\psi$ does not contribute to the derivatives of the equilibrium free energy. Hence, the reduced expression (\ref{F short}) is sufficient to determine the true minimum.

For a helical state with $q \neq 0$, minimization with respect to $q^2$ provides the relation
\begin{equation}
    \zeta\sin^2\psi = A[\sin^2(\psi-\eta)].
    \label{dF short dq}
\end{equation}
Similarly, differentiating the free energy with respect to $\psi$ and using Eq.\ (\ref{dF short dq}), we obtain
\begin{equation}
    \frac{dF}{d\psi}=-\frac{q^2}{\sin\psi}A[\sin(\psi-\eta)\sin\eta]\leq0.
\end{equation}
This expression is negative since the triplet vector $\mathbf{m}$ always lies between $\mathbf{h}$ and $\mathbf{\hat{a}}$, see Fig.\ \ref{fig:cone}. We conclude that the minimum of the free energy with respect to $\psi$ is achieved at the highest possible value of this angle $\psi = \pi/2$. The optimal helical configuration is thus the planar spin spiral, in which the magnetization rotates in a single plane perpendicular to the rotation axis $\mathbf{\hat{a}}$.

\bibliography{bibliography.bib}
\end{document}